\documentclass[letterpaper]{article} % DO NOT CHANGE THIS
\usepackage{aaai25}  % DO NOT CHANGE THIS
\nocopyright
\usepackage{times}  % DO NOT CHANGE THIS
\usepackage{helvet}  % DO NOT CHANGE THIS
\usepackage{courier}  % DO NOT CHANGE THIS
\usepackage[hyphens]{url}  % DO NOT CHANGE THIS
\usepackage{graphicx} % DO NOT CHANGE THIS
\usepackage{natbib}  % DO NOT CHANGE THIS AND DO NOT ADD ANY OPTIONS TO IT
\usepackage{caption} % DO NOT CHANGE THIS AND DO NOT ADD ANY OPTIONS TO IT
\usepackage{algorithm}
\usepackage{color}
\usepackage{algorithmic}
\usepackage{utfsym}

\usepackage{newfloat}
\usepackage{listings}
\DeclareCaptionStyle{ruled}{labelfont=normalfont,labelsep=colon,strut=off} % DO NOT CHANGE THIS
\floatstyle{ruled}
\newfloat{listing}{tb}{lst}{}
\floatname{listing}{Listing}
\title{Drop the beat! Freestyler for Accompaniment Conditioned \\Rapping Voice Generation}
\author{
    Ziqian Ning\textsuperscript{\rm 1,2}, Shuai Wang\textsuperscript{\rm 3}, Yuepeng Jiang\textsuperscript{\rm 1}, Jixun Yao\textsuperscript{\rm 1}, \\Lei He\textsuperscript{\rm 2}, Shifeng Pan\textsuperscript{\rm 2}, Jie Ding\textsuperscript{\rm 2}, Lei Xie\textsuperscript{\rm 1}\thanks{Corresponding Author.}
}
\affiliations{
    \textsuperscript{\rm 1}Audio, Speech and Language Processing Group (ASLP@NPU), School of Computer Science, \\ Northwestern Polytechnical University, Xi'an, China\\
    \textsuperscript{\rm 2} Microsoft, China\\
    \textsuperscript{\rm 3} Shenzhen Research Institute of Big Data, \\The Chinese University of Hong Kong, Shenzhen (CUHK-Shenzhen), China\\
    \{ningziqian, Jiangyp, yaojx\}@mail.nwpu.edu.cn, wangshuai@cuhk.edu.cn,\\
    \{helei, peterpan, jie.DING\}@microsoft.com, lxie.nwpu.edu.cn
}

\usepackage{bibentry}
\usepackage{subfig}
\usepackage{amssymb,amsmath}
\usepackage{bbding}
\usepackage{multirow}

\usepackage{cuted}
\begin{document}

\maketitle

\begin{abstract}
Rap, a prominent genre of vocal performance, remains underexplored in vocal generation. General vocal synthesis depends on precise note and duration inputs, requiring users to have related musical knowledge, which limits flexibility. In contrast, rap typically features simpler melodies, with a core focus on a strong rhythmic sense that harmonizes with accompanying beats. In this paper, we propose \textit{Freestyler}, the first system that generates rapping vocals directly from lyrics and accompaniment inputs. Freestyler utilizes language model-based token generation, followed by a conditional flow matching model to produce spectrograms and a neural vocoder to restore audio. It allows a 3-second prompt to enable zero-shot timbre control. Due to the scarcity of publicly available rap datasets, we also present \textit{RapBank}, a rap song dataset collected from the internet, alongside a meticulously designed processing pipeline.
%we collected a large volume of rap songs from the internet and designed a meticulous pipeline for data cleaning, processing, and filtering.
% \footnote{The dataset and processing pipeline will be released}. 
Experimental results show that Freestyler produces high-quality rapping voice generation with enhanced naturalness and strong alignment with accompanying beats, both stylistically and rhythmically.%}
\end{abstract}


\section{Introduction}

%As one of the most unique genre of singing, rap has neither been specifically been studied in SVS nor TTSong. The core characteristics of rap music are rhythm and tempo, and what makes it distinctly different from other genres.Rap music focuses heavily on rhythmic and beat accuracy. Often, rap artists will rap quickly in a powerful and tight manner over a background beat. This tight combination makes rap music sound very dynamic and full of energy. To generate raps that can be extremely rhythmic, a natural idea is to reverse the generation order and synthesis vocal condition on the accompaniment.
Rap stands out as one of the most distinctive genres of vocal performance, yet it has received limited attention in the field of vocal generation. At its core, rap is defined by its emphasis on rhythm and tempo, distinguishing it markedly from other genres. Rappers typically deliver rapid, powerful verses that tightly synchronize with the accompanying beats, creating a dynamic and energetic auditory experience. 

\begin{figure}[ht]
\centering
\includegraphics[scale=0.60]{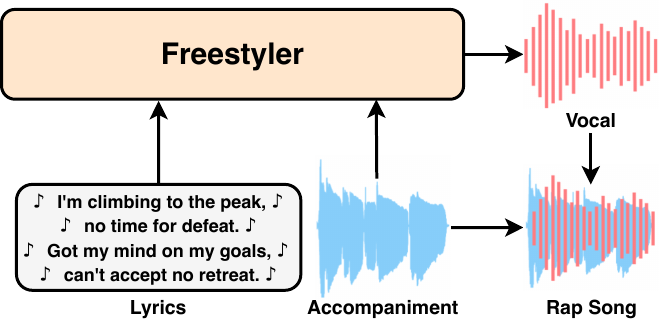}
\vspace{-4pt}
\caption{The overall pipeline of Freestyler. With lyrics and accompaniment as condition, it can generate rapping voice that matches the style and rhythm of the accompaniment.}
\label{fig:lm}
\vspace{-12pt}
\end{figure}
%Singing voice synthesis (SVS) requires lyrics along with notes and durations as input. Its predefined rhymes do not lend themselves to the expressive flexibility needed for genres like rap. Text-to-song (TTSong), on the other hand, can generate vocals and accompaniment based on lyrics and style prompts. A common approach for TTSong involves generating the vocal track by utilizing the lyrics as input, followed by the creation of the accompaniment track through the use of natural language prompts and the generated vocals. Nevertheless, when it comes to rapping voice generation, generating the vocals before the accompaniment can cause the vocals to be lack of rhythm. %To produce rhythmic raps, it is better to reverse the generation order and condition the vocal synthesis on the accompaniment.
Normal singing voice synthesis (SVS) requires lyrics, notes, and duration as inputs. Given that rap is inherently a freeform performance style characterized by varied rhythms, predefined rhythms in SVS can hinder the naturalness of the generation process. In contrast, Text-to-song (TTSong) offers greater flexibility, generating both vocals and accompaniment solely from lyrics and style prompts. A typical TTSong approach involves first creating the vocal track based on the lyrics and then generating the accompaniment track using natural language prompts alongside the produced vocals. However, generating rap vocals with only lyrics condition may compromise the rhythmic integrity.
Furthermore, the limited availability of rap datasets poses an additional challenge in rapping voice generation. Singing data is much less abundant compared to large-scale speech datasets, and within this limited pool, rap data is an even smaller subset. This scarcity of rap data further complicates the process of generating rapping voices.

In this paper, we present Freestyler, the first rapping voice generation model capable of generating rap that harmonizes with the style and rhythm of the accompaniment using only lyrics and accompaniment inputs. With a 3-second reference audio, it can adapt to any speaker's voice. Our approach employs a three-stage framework: lyrics-to-semantic, semantic-to-spectrogram, and spectrogram-to-audio. To address the challenge of data scarcity, we use discrete semantic tokens as a proxy representation so that some parts of the model do not require supervised data for training. The first stage adopts a language model to predict discrete semantic tokens conditioned on lyrics and fine-grained accompaniment features. The second stage applies conditional flow matching techniques for mel-spectrogram prediction. Finally, a vocoder restores audio from the spectrogram. Given the lack of publicly available rap datasets, we collected a large volume of rap songs from the internet and designed a meticulous pipeline for data cleaning, processing, and filtering, resulting in a dataset we have named RapBank.

Experiments conducted on our collected rap dataset show that Freestyler generates high-quality rap that fits the accompaniment. The main contributions of this work are summarized as follows:
\begin{itemize}
\item We propose Freestyler, the first accompanied rapping voice generation model that takes lyrics and accompanying music as conditions.
\item %We collected a large amount of rap data and designed a comprehensive data-processing pipeline, making it suitable for training. 
We present RapBank, a large volume rap dataset with comprehensive data-processing pipeline, suitable for model training. Both the data and the processing pipeline are publically available on Github\footnote{https://github.com/NZqian/RapBank}.%will be open source.
\item We developed a language model that uses accompaniment conditions to provide global style control and fine-grained rhythm control of vocal production. We also adopt conditional flow matching for high-quality mel-spectrogram prediction.
\item Experimental results from objective and subjective evaluations demonstrate the effectiveness of Freestyler\footnote{Samples: https://nzqian.github.io/Freestyler/}.
\end{itemize}

\section{Related work}
\subsection{Singing Voice Synthesis}
Singing voice synthesis (SVS) aims to generate natural singing voices based on lyrics, musical scores, and corresponding durations. VISinger 2~\cite{visinger2} introduces an end-to-end system utilizing a digital signal processing (DSP) synthesizer to enhance sound quality. NaturalSpeech 2~\cite{naturalspeech2} and StyleSinger~\cite{stylesinger} employ a reference voice clip for timbre and style extraction, enabling style transfer and zero-shot synthesis. PromptSinger~\cite{promptsinger} is the first system to attempt guiding singing voice generation through text descriptions, placing greater emphasis on speaker identity and timbre control. DiffSinger~\cite{diffsinger} addresses the issue of excessive smoothness by implementing a shallow diffusion mechanism. To bridge the gap between realistic music scores and detailed MIDI annotations, RMSSinger~\cite{rmssinger} proposes a word-level modeling approach combined with diffusion-based pitch prediction. MIDI-Voice~\cite{midivoice} incorporates MIDI-based priors for expressive zero-shot generation. VoiceTuner~\cite{voicetuner} advocates a self-supervised pre-training and fine-tuning strategy to mitigate data scarcity, applicable to low-resource SVS tasks. Despite contributions from open-source singing voice datasets~\cite{opencpop,m4singer,nus48e}, their quantity significantly lags behind that of speech datasets, and none specifically cater to rap genres.
\subsection{Music Generation}
%Music generation is a general task that includes symbolic music generation, waveform generation, and accompaniment generation. MuseGAN~\cite{musegan} achieves symbolic music generation via a GAN-based approach. SongMASS~\cite{songmass} designs a songwriting method that generates lyrics or melodies, conditioned on each other. SongComposer\cite{songcomposer} proposes a music large language model (LLM) for song composition, which can compose melodies and lyrics with symbolic song representations. Inspired by the two-stage modeling in audio generation~\cite{audiolm}, MusicLM~\cite{musiclm} adopts a cascade of transformer decoders to sequentially generate the semantic tokens and the acoustic tokens, conditioned on joint textual-music representations from MuLan~\cite{mulan}. MusicGen~\cite{musicgen} proposes a novel codebook interleaving patterns to generate music codec tokens in a single transformer decoder, while MeLoDy~\cite{melody} introduces an LM-guided diffusion model that efficiently generates music audios. MusicLDM~\cite{musicldm} incorporates beat-tracking information and applies data augmentation through latent mixup to address the potential plagiarism issue in music generation. However, music generation models are not required to produce perceptible singing voices, which is still a challenge. In addition, several works focus on singing-to-accompaniment generation. SingSong~\cite{singsong} is proposed to generate instrumental music to accompany input vocals. Melodist~\cite{melodist} incorporates a transformer decoder to achieve controllable accompaniment generation.

Music generation encompasses various tasks, including symbolic music generation, lyrics generation, and accompaniment generation. MuseGAN~\cite{musegan} achieves symbolic music generation through a GAN-based approach. SongMASS~\cite{songmass} designs a method for songwriting that generates lyrics or melodies conditioned on each other, while SongComposer~\cite{songcomposer} proposes a large language model (LLM) for song composition, capable of generating melodies and lyrics with symbolic song representations. DeepRapper~\cite{deeprapper} focuses on rap lyrics generation, which also leverages an LLM to generate lyrics from right to left with rhyme constraints.
Inspired by two-stage modeling in audio generation~\cite{audiolm}, MusicLM~\cite{musiclm} uses a cascade of transformer decoders to sequentially generate semantic and acoustic tokens, based on joint textual-music representations from MuLan~\cite{mulan}. MusicGen~\cite{musicgen} introduces a novel approach with codebook interleaving patterns to generate music codec tokens in a single transformer decoder, which is further combined with stack patterns in ~\citealp{stack-and-delay} to improve generation quality. Additionally, MeLoDy~\cite{melody} presents an LM-guided diffusion model that efficiently generates music audio, and MusicLDM~\cite{musicldm} incorporates beat-tracking information and latent mixup data augmentation to address potential plagiarism issues in music generation.
Several works focus specifically on vocal-to-accompaniment generation, such as SingSong~\cite{singsong}, which generates instrumental music to accompany input vocals, and Melodist~\cite{melodist}, which utilizes a transformer decoder for controllable accompaniment generation.

\subsection{Text-to-Song}
%Text-to-song (TTSong), also known as accompanied singing voice synthesis (ASVS), aims to produce natural singing voices accompanied by music. TSong incorporates elements from both music generation and singing voice synthesis; the former primarily focuses on music creation, while the latter is concerned with singing voice synthesis.
%A prevalent method in TTSong involves a two-stage approach: first, generating the vocal track based on lyrics input, followed by predicting accompaniment that complements the vocal output. Melodist~\cite{melodist} is the first TTSong model, which adopts two autoregressive transformers to sequentially generate vocal codec tokens and accompaniment codec tokens conditioned on lyrics, musical scores and natural language prompts. MelodyLM~\cite{melodylm} eliminates the need for users to input music scores by relying solely on textual descriptions and vocal references as conditions.
Text-to-song (TTSong), also recognized as Accompanied Singing Voice Synthesis (ASVS), strives to produce natural singing voices accompanied by music. TTSong incorporates elements from both singing voice synthesis and music generation; the former focuses on generating a singing vocal, while the latter primarily involves music creation. A common methodology in TTSong employs a two-stage process: initially generating the vocal track from lyrical input, followed by the prediction of accompanying music. Melodist~\cite{melodist} is the first TTSong model, which utilizes two autoregressive transformers to sequentially produce vocal and accompaniment codec tokens, conditioned on lyrics, musical scores, and natural language prompts. 
MelodyLM~\cite{melodylm} eliminates the need for music scores in Melodist and instead relies solely on textual descriptions and vocal references.

% \begin{figure*}[!htbp]
% \centering
% \includegraphics[scale=0.65]{figs/overall.drawio.pdf}

% \caption{The architecture of DualVC 3. Implicit knowledge is achieved by applying a dynamic chunk mask in Conformer blocks, allowing dual-mode inference that can accept both full sequence and chunked speech as input. HPC and Wav2Vec 2.0 are discarded after training.}
% \label{fig:overall}
% \end{figure*}

\begin{figure*}[!htbp]
  \begin{minipage}[t]{0.5\linewidth}
    \centering    
\subfloat[Lyrics-to-Semantic]{\includegraphics[clip,scale=0.55]{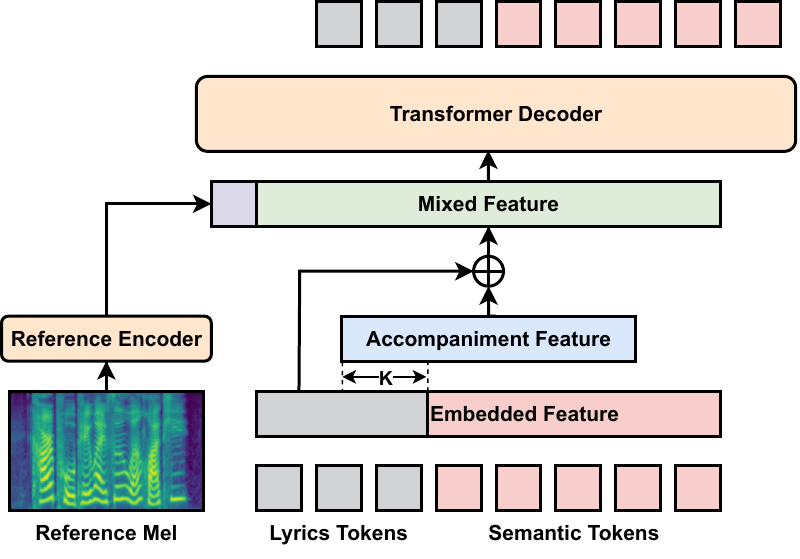}
    \label{fig:a}}
  \end{minipage}%
  \begin{minipage}[t]{0.5\linewidth}
    \centering
\subfloat[Semantic-to-Spectrogram]{\includegraphics[clip,scale=0.55]{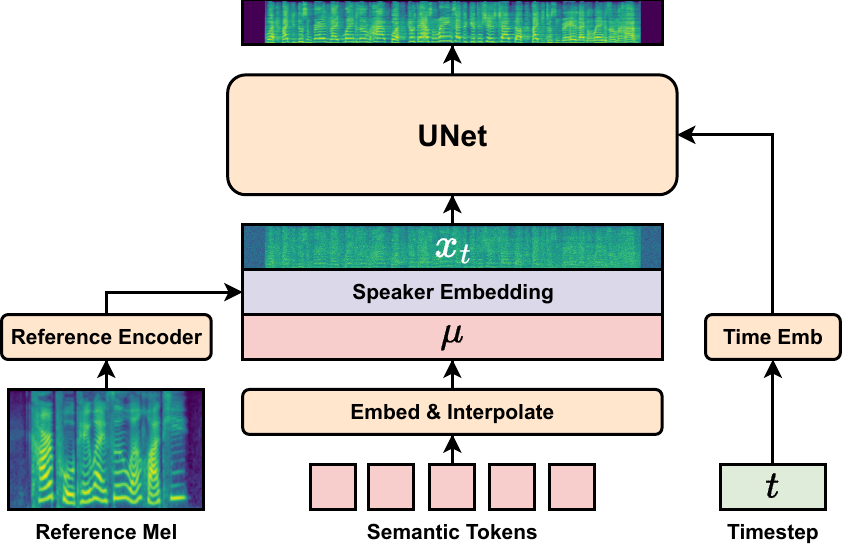} % semantic-to-spectrogram
\label{fig:b}}
  \end{minipage}
     \caption{Overview of Freestyler. The lyrics-to-semantic model in (a) predicts semantic tokens based on lyrics and accompaniment. The accompaniment feature is shifted left by $K$ frames to provide additional rhythmic context. The semantic-to-spectrogram model in (b) generates mel-spectrograms from the semantic tokens, which are interpolated to align with the spectrogram's frame rate. Speaker embedding is provided to both models to control the timbre.}
    \label{fig:overall}
    \vspace{-12pt}
\end{figure*}

\section{Freestyler}

In this section, we first define the task of accompaniment-conditioned rapping voice generation. Subsequently, we provide an overview of the proposed system, Freestyler, followed by a detailed explanation of each stage within it.

\subsection{Task Definition}
In this study, we present a novel task: \textit{rapping voice generation}. This task entails the creation of rapping voices that are stylistically and rhythmically synchronized with the accompanying music. It can be regarded as the inverse of accompaniment generation~\cite{singsong}, which generates instrumental music to accompany input vocals
%Given the training set $S$ consists of $n$ songs, 

Given the training dataset consisting of rap songs $R$ and their corresponding lyrics $L$, we separate $R$ into the vocal tracks $R_v$ and the accompaniment tracks $R_a$. A short segment $C$ is randomly extracted from $R_v$ as the reference audio which presents the rapper's timbre. The task of \textit{rapping voice generation} can be defined as modeling the conditional probability distribution $p(R_v|R_a, L, C)$.

\subsection{Overview}
To address the challenge of data scarcity, we divide the task of rapping voice generation into three hierarchical stages: lyrics-to-semantic, semantic-to-spectrogram, and spectrogram-to-audio. Rather than directly generating acoustic tokens or mel-spectrograms from lyrics, we employ semantic tokens derived from K-means clustering on self-supervised learning (SSL) representations~\cite{speartts,promptvc} as a proxy feature to bridge the lyrics-to-semantic and semantic-to-spectrogram stages. This method offers two primary advantages. First, semantic tokens are more closely aligned with the text domain, enabling the first-stage model to be trained using less annotated data. Second, the subsequent two stages can be trained in an unsupervised manner by utilizing more unlabeled data. We use a language model (LM) for the lyrics-to-semantic stage conditioned on lyrics, accompaniment features, and a 3-second reference audio segment, which generates discrete semantic tokens. For the semantic-to-spectrogram stage, we employ a conditional flow matching (CFM) model to transform the discrete semantic tokens into continuous mel-spectrograms. The reference audio is also incorporated into the CFM model to complement the missing timbre information contained in the semantic tokens. Finally, a pre-trained vocoder is utilized to reconstruct audio from the spectrogram. The overall model design is illustrated in Figure~\ref{fig:overall}.

%\subsection{Fine-grained Accompaniment Conditioned Language Model for lyrics-to-semantic}
\subsection{Lyrics-to-Semantic}
\begin{figure}[ht]
\centering
\includegraphics[scale=0.70]{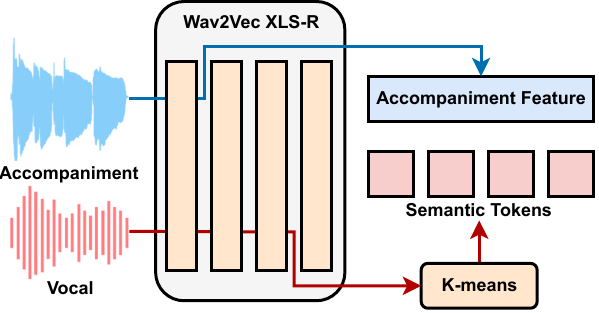}

\caption{The extraction process of the accompaniment feature and semantic tokens. Each block in Wav2Vec XLS-R represents 6 attention layers, with accompaniment and vocals going through 6 and 18 layers respectively.}
\label{fig:feat}
\vspace{-12pt}
\end{figure}
\subsubsection{Feature Representation and Tokenization}
As discussed in~\citealp{w2vlayer, wavlm}, different layers of SSL models encode different types or extents of information, with shallower layers capturing more acoustic features and deeper layers representing more semantic aspects. Therefore, with paired vocal and accompaniment as inputs, we extract discrete semantic tokens and continuous accompaniment features using two different layers of an SSL model. As shown in Figure~\ref{fig:feat}, we utilize a pre-trained Wav2Vec XLS-R~\cite{xlsr} for feature extraction, where the vocal input is processed through its 18 layers, and the semantic tokens $S$ are subsequently obtained using K-means clustering. While the accompaniment input is passed through 6 layers to derive the accompaniment features $A$. For the lyrics, we employ a grapheme-to-phoneme (G2P) phonemizer~\cite{g2p} to obtain the lyrics tokens $L$.

\subsubsection{Language Model with Fine-grained Accompaniment  Condition}
\label{sec:llm_con}
% As illustrated in Figure~\ref{fig:a}, Freestyler employs an autoregressive transformer decoder to generate the semantic tokens in a language modeling manner. The conditions include lyrics, accompaniment features, and a reference mel-spectrogram.

% To achieve global style control and make the generated rap synchronize with the accompaniment, we incorporated fine-grained accompaniment features as conditions. Initially, the lyrics tokens and semantic tokens are concatenated and subsequently embedded, which are then summed with the accompaniment features to produce the mixed feature.
% The generation process of an autoregressive language model is a unidirectional next-token-prediction task. The model has only historical receptive fields without being able to see future information. If the accompaniment condition is aligned with semantic tokens on the time axis, the model does not have enough stylistic and rhythmic conditions at the beginning of the inference process, making the rap to be less matched to the accompaniment, and will accumulate this error in the autoregressive process. Therefore, we shift the accompaniment feature left by $K$ frames, so that the model can have the accompaniment information of $K+t$ frames when generating the $t$-th frame. Thus we model the following distribution:

As illustrated in Figure~\ref{fig:a}, the generative process of semantic tokens in Freestyler is formulated as a unidirectional next-token-prediction task, i.e., in the form of a language model. In the prediction process, Freestyler relies on lyrics, accompaniment features, and a reference mel-spectrogram to constrain the generated semantic tokens. More specifically, initially, the lyrics tokens and semantic tokens are concatenated and subsequently embedded, which are then summed with the accompaniment features to produce the mixed feature. However, based on our experience, if the accompaniment condition is exactly aligned with the semantic tokens, the generated rap vocal and accompaniment will have a certain degree of mismatch. We hypothesize this is because the context information of the accompaniment is quite important, but due to the current model design which has no access to future information, such context information cannot be effectively learned. Based on this, we shift the accompaniment feature left by $K$ frames, so that the model can have the accompaniment information of $K+t$ frames when generating the $t$-th frame. Thus we model the following distribution:

\begin{equation}
p(S) = \prod^{n}_{t=2}p(s_t|s_{<t},l, c, a_{<t+K};\theta_{LM}),
\end{equation}
where $s$, $l$, $c$, and $a$ stand for semantic tokens, lyrics tokens, speaker embedding extracted using a reference encoder, and acoustic features.

During training, the vocal-accompaniment pairs are of equal length. However, during inference, the lengths of these two elements may vary. This mismatch can lead to early termination of the model if the accompaniment is too short, or result in the generation of hallucinated content if the accompaniment is excessively long. To address this issue, we introduce a random mask on the accompaniment condition, thereby mitigating the strong temporal correlation between these two features.

\subsubsection{Zero-shot Timbre Control}
% The predominant zero-shot approaches for TTS or SVS involve either to use prompts for timbre control leveraging the language model's in-context learning capabilities, or to use a speaker embedding as global timbre condition. However, for accompaniment-conditioned rapping voice generation, using prompts for timbre control would require a piece of the target speaker's rap data and corresponding accompaniment feature. This requirement imposes significant constraints on the inference process. To address this limitation, we propose the use of a reference encoder to extract global speaker embedding from a segment of reference audio, concatenated to the head of the mixed features for timbre control. This also enables a normal speaker to sing rap.

The predominant zero-shot approaches for TTS or SVS usually involve two main methods: using voice prompts to leverage the language model's in-context learning capabilities, or using a speaker embedding as a global timbre condition. In Freestyler, the latter approach was adopted. 
The rationale behind this choice is that the style of the voice prompt can significantly influence the final generated speech style. This means we would need to require the user to provide a rapping segment as the prompt, which is non-trivial.

On the other hand, the speaker embedding mainly affects the timbre, while the rapping style is primarily controlled by the accompaniment features. This ensures we can generate rapping vocals with any target timbre.
In the specific implementation, we propose the use of a reference encoder to extract a global speaker embedding from a segment of reference audio. This embedding is then concatenated to the head of the mixed features for timbre control.

\subsection{Semantic-to-Spectrogram}
We employ conditional flow matching to map semantic tokens to mel-spectrograms. Given that semantic tokens lack complete timbre information due to their discrete nature, we incorporate additional timbre conditions at this stage. 
Let $\boldsymbol{x}$ denote an observation in the data space $\mathbb{R}^d$, sampled from a complicated, unknown data distribution $q(\boldsymbol{x})$. 
A probability density path is a time-dependent probability density function, $p_t : [0, 1] \times \mathbb{R}^d \rightarrow \mathbb{R} > 0$. 
One way to generate samples from the data distribution $q$ is to construct a probability density path $p_t$, where $t \in [0, 1]$ and $p_0(\boldsymbol{x}) = \mathcal{N} (\boldsymbol{x}; \boldsymbol{0}, \boldsymbol{I})$ is a prior distribution, such that $p_1(\boldsymbol{x})$ approximates the data distribution $q(\boldsymbol{x})$. 
For example, CNFs first define a vector field $\boldsymbol{v_t} : [0, 1] \times \mathbb{R}^d \rightarrow \mathbb{R}^d$, which generates the flow $\phi_t : [0, 1] \times \mathbb{R}^d \rightarrow \mathbb{R}^d$ through the ODE

\begin{equation}
\frac{d}{dt}\phi_t(\boldsymbol{x}) = \boldsymbol{v}_t(\phi_t(\boldsymbol{x})); \qquad \phi_0(\boldsymbol{x}) = \boldsymbol{x}.
\label{eq:ode}
\end{equation}
This generates the path $p_t$ as the marginal probability distribution of the data points. We can sample from the approximated data distribution $p_1$ by solving the initial value problem in Eq.~\ref{eq:ode}.

Suppose there exists a known vector field $\boldsymbol{u_t}$ that generates a probability path $p_t$ from $p_0$ to $p_1 \approx q$, conditional flow matching considers

\begin{equation}
\mathcal{L}_{CFM}= \mathbb{E}_{t,q(\boldsymbol{x_1}),p_t(\boldsymbol{x}|\boldsymbol{x_1})}\Vert\boldsymbol{u_t}(\boldsymbol{x}|\boldsymbol{x_1})-\boldsymbol{v_t}(\boldsymbol{x}|\mu, \hat{c};\theta)\Vert^2.
\label{eq:cfm}
\end{equation}

where $t \sim \mathbb{U}[0, 1]$ and $\boldsymbol{v_t}(\boldsymbol{x}|\mu, \hat{c};\theta)$ is a neural network with parameters $\theta$. $\mu$ is the embedded and interpolated semantic tokens and $\hat{c}$ is the timbre embedding. This replaces the intractable marginal probability densities and the vector field in flow matching loss with conditional probability densities and conditional vector fields.

% The semantic-to-mel model is trained using optimal-transport conditional flow matching (OT-CFM) [14], which is a CFM variant with particularly simple gradients. The OT-CFM loss function can be written

% \begin{equation}
% \mathcal{L}(\theta) = \mathbb{E}_{t,q(\boldsymbol{x_1}),p_0(\boldsymbol{x_1})}\Vert\boldsymbol{u_t}^{OT}(\phi_t^{OT}(\boldsymbol{x})|\boldsymbol{x_1})-\boldsymbol{v_t}(\phi_t^{OT}(\boldsymbol{x})|\boldsymbol{\mu};\theta)\Vert^2,
% \label{eq:ot-cfm}
% \end{equation}

% defining $\phi_t^{OT}(\boldsymbol{x}) = (1 - (1 - \sigma_{min})t)\boldsymbol{x_0} + t\boldsymbol{x_1}$ as the flow from $\boldsymbol{x_0}$ to $\boldsymbol{x_1}$ where each datum $\boldsymbol{x_1}$ is matched to a random sample $\boldsymbol{x_0} \sim \mathcal{N} (\bold{0}, \boldsymbol{I})$ as in [14]. Its gradient vector field – whose expected value is the target for the learning is then $\boldsymbol{u_t}^{OT}(\phi_t^{OT}(\boldsymbol{x})|\boldsymbol{x_1}) = \boldsymbol{x_1} - (1- \sigma_{min})\boldsymbol{x_0}$, which is linear, time-invariant, and only depends on $\boldsymbol{x_0}$ and $\boldsymbol{x_1}$. These properties enable easier and faster training, faster generation, and better performance compared to DPMs.

As shown in Figure~\ref{fig:b}, the conditional flow matching model follows Matcha-TTS~\cite{matchatts} to use U-Net~\cite{unet} architecture as the backbone, containing 1D convolutional residual blocks to downsample and upsample the inputs, with the flow matching step $t \in [0, 1]$ embedded as in ~\citealp{gradtts}. Each residual block is followed by a Transformer block, whose feedforward nets use snake beta activations~\cite{bigvgan}.  Additionally, the speaker embedding $\hat{c}$ is extracted by a reference encoder. Note that the reference encoder in the LM and CFM do not share parameters.

\subsection{Spectrogram-to-audio}
We employ the V2 version of BigVGAN~\cite{bigvgan} for audio restoration. Compared to the V1 version, BigVGAN-V2 is trained using datasets containing diverse audio types, including speech in multiple languages, singing, and environmental sounds. Also, the discriminator is improved with multi-scale sub-band CQT discriminator~\cite{mssbcqtdisc} and multi-scale mel-spectogram loss~\cite{msmelloss}. We found BigVGAN-V2 exhibits high sound quality and exceptional robustness on rapping voice.

\section{RapBank}
%Given that no rap dataset is available, we construct and release a rap dataset called RapBank, which consists of songs, lyrics, and multiple quality-related metrics. 
% In light of the absence of a dedicated rap dataset, we developed and released the first rap dataset named RapBank.

To the best of our knowledge, there is currently no publicly available dataset for rapping synthesis. Thus, we created an automatic pipeline to collect and label a novel dataset, RapBank, for the proposed rapping synthesis task.

RapBank comprises $92,371$ rap songs with a total duration of $5,586$ hours. After segmentation, we produced $904,548$ rap clips with an duration of $4,353$ hours, with corresponding lyrics and various quality-related metrics. The English subset utilized to train the LM will be presented in the experimental section, while detailed statistical information about RapBank can be found in the appendix.

\subsubsection{Data Crawling}
Initially, we crawled all available rap-related playlists on YouTube using the keywords ``Rap" and ``Hip-hop". Next, we extracted distinct video IDs from the playlists and filtered out those with a duration exceeding ten minutes, as they are unlikely to be songs. Then we downloaded the videos using their respective IDs and retain only the audio track. We resampled the audios to 44.1 kHz, and average stereo mixes to mono. %Additionally, we downloaded the corresponding lyrics where available.

\subsubsection{Source Seperation}
In the previous step, we collected songs with both vocals and accompaniments mixed in a single track. As we aim to generate rapping vocals with accompaniment and lyrics as conditions, the first stage of data processing involves separating vocals and accompaniment into different tracks. To accomplish this, we utilized the state-of-the-art music source separation model, BS-RoFormer~\cite{bsroformer}. We first extract the vocals from the mixed track and subsequently subtract them from the original track to obtain the pure accompaniment. %Considering that songs typically last three to four minutes, processing them directly with the non-streaming BS-RoFormer model can be impractical due to limited GPU memory, We employ a chunk-wise inference approach during the separation process. Specifically, we extract 8s clips with 4s overlap from each song and input resultant clips to the BS-RoFormer. The outputs are concatenated to produce the separated vocals and accompaniment.

\subsubsection{Segmentation}
To ensure the data length is suitable for model training while eliminating non-speech segments, we employed Voice Activity Detection (VAD) to segment the separated vocal tracks. We utilized WebRTC Voice Activity Detector~\footnote{https://github.com/wiseman/py-webrtcvad} to extract frame-level voice/unvoice labels and subsequently slice the vocal track into segments containing only vocal sounds. Adjacent segments are sequentially merged if the unvoice gap between them is less than three seconds, continuing this process until the total duration exceeds a specified threshold. This threshold is sampled from a Gaussian distribution with a mean of 18 seconds to enhance variability in data lengths. Subsequently, We sliced the accompaniment with the same timestamp to obtain segments that correspond the vocals.

\subsubsection{Lyrics Recognition}
%Less than one-tenth of the songs we crawled contain lyrics. With unclean textual content and inaccurate timestamps, they are not suitable for model training. Thus we designed a lyrics recognition stage using Whisper~\cite{whisper} to transcribe lyrics automatically. Whisper is an automatic speech recognition (ASR) model trained on an extensive dataset comprising 680,000 hours of multilingual speech. It demonstrates high accuracy and robustness across various languages. Given the absence of ASR models specifically designed for lyrics recognition, Whisper is the most powerful ASR model available. It is noteworthy that Whisper exhibits a higher word error rate (WER) when transcribing rap songs compared to speech data. This increase in WER is to be expected, as rap features a rapid delivery, heightened expressiveness, and frequent use of alliteration. 

%Most recognition errors can be classified into two categories. The first category involves misrecognition resulting from words with similar pronunciations, which is unlikely to affect model training. The second category comprises hallucinations, where the model produces phrases or sentences that do not exist in any form within the underlying audio. Hallucinations often lead to an abnormal speaking rate; therefore, we can filter these results based on the speaking rate, a method that will be discussed in the following section.

% Less than one-tenth of the songs we crawled contain lyrics. With unclean textual content and inaccurate timestamps, they are not suitable for model training.
Another challenge we faced in designing RapBank was that less than one-tenth of the songs we collected contained lyrics. Additionally, issues such as unclean textual content and inaccurate timestamps made these data difficult to be used directly for model training.
Thus we employ the powerful automatic speech recognition (ASR) model Whisper~\cite{whisper} to transcribe the lyrics. As Whisper is trained with speech data, the word error rate (WER) of resulting lyrics is significantly higher than speech. The recognition errors can be categorized into two types. The first type involves misrecognition resulting from words with similar pronunciations, which has minimal impact on model training. The second type involves hallucinations, where the model produces phrases or sentences that do not exist in any form within the input audio. Hallucinations often lead to an abnormal singing tempo; therefore, we can filter these results based on the singing tempo, which will be discussed in the following section.

\subsubsection{Quality Filtering}
%Due to the common occurrence of digital effects in songs and multi-singer scenearios, we further employ a series of filtering methods to ensure the quality of the final processed rap data. Specifically, We use a speaker diarization system to calculate the speech duration of each speaker and identify the person with the longest speaking duration, named the primary speaker. Then we discard all sentences where the primary speaker's duration is less than 80\% of the total sentence duration. Then we extract the phoneme from the transcripts and calculated the number of phoneme-per-second (PPS), discarding segments with PPS less than 12 and greater than 30. Most of the transcripts corresponding to the abnormal PPS segments contained hallucinations. 
Given the prevalence of digital effects and multi-singer scenarios in songs, we adopt several quality-related metrics apply filtering to ensure the quality of the final processed rap data. 
Following~\citealp{wenetspeech4tts}, we further devide the dataset into three subsets with increasing quality—Basic, Standard, and Premium—utilizing these metrics for multi-stage training.
Specifically, we utilize a diarization model~\footnote{https://github.com/pyannote/pyannote-audio} to calculate the singing duration for each singer and identify the individual with the longest singing duration as the primary singer. 
% Subsequently, we exclude all segments where the primary singer's duration accounts for less than 80\% of the total duration. 
Subsequently, we exclude all segments where the primary singer's duration does not occupy the majority, i.e., less than 80\% (can vary for different subsets) of the total duration.
We extract phonemes from the lyrics and compute the phoneme-per-second (PPS) rate, which reflects the tempo of the singing voice. Segments with a PPS rate that are either too low or too high are discarded, as most lyrics associated with these segments exhibit hallucinations.
%We observed a significantly higher incidence of hallucinations in languages other than English, probably due to the fact that other languages make up a smaller percentage of the whisper training data compared to English. Therefore in this paper, we filtered out all songs other than English.
To eliminate low-quality speech segments, we use the DNSMOS P.808 scores~\cite{dnsmos} to evaluate each rap segment. %Following ~\citealp{wenetspeech4tts}, we devide the dataset into three subsets with increasing quality—Basic, Standard, and Premium—utilizing the aforementioned metrics for multi-stage training. %These subsets contains 2000, 500, 100 hours of data, respectively.
%We calculate the duration of individual speakers using a speaker diarization model, and discard segments in which the speaker with the longest duration accounted for less than 80\% of the duration of the entire sentence.
\section{Experimental Setup}
\label{sec:exp}

\begin{table*}[!htbp]

\caption{Objective and subjective evaluation of vocals generated by Freestyler, including different configurations of whether accompaniment conditions are used in the training and inference stage}
\vspace{-6pt}
 \label{tab:vocal}
 \centering
 %\resizebox{\linewidth}{!}{
% \setlength{\tabcolsep}{4pt}
\renewcommand{\arraystretch}{1.2}
\begin{tabular}{ccccccccc}
%\hline
%Model      & Training w/ Acco & Inference w/ Acco & MOS-N$\uparrow$ & MOS-R$\uparrow$ & WER$\downarrow$  \\ \hline
%GT Vocal   &       -           &        -           & 4.27$\pm0.06$   & 4.13$\pm0.05$   &  31.7            \\ \hline
%\multirow{3}{*}{Freestyler} & \usym{2717}      &    \usym{2717}    & 3.43$\pm0.03$   & 3.21$\pm0.07$   &  32.6            \\
% & \usym{2713}        &    \usym{2717}    & 3.68$\pm0.04$   & 3.55$\pm0.06$   &  32.4            \\
% & \usym{2713}        &    \usym{2713}      & 3.92$\pm0.05$   & 3.80$\pm0.06$   &  33.2            \\ \hline
\hline
Model                       & Training w/ Acco         & Inference w/ Acco        & MOS-N$\uparrow$ & MOS-R$\uparrow$ & WER$\downarrow$ \\ \hline
GT Vocal                    & -                        & -                        & 4.27$\pm0.06$   & 4.13$\pm0.05$   & 31.7            \\ \hline
\multirow{3}{*}{Freestyler} & \usym{2717} & \usym{2717} & 3.43$\pm0.03$   & 3.21$\pm0.07$   & 32.6            \\
                            & \usym{2713} & \usym{2717} & 3.68$\pm0.04$   & 3.55$\pm0.06$   & 32.4            \\
                            & \usym{2713} & \usym{2713} & 3.92$\pm0.05$   & 3.80$\pm0.06$   & 33.2            \\ \hline
\end{tabular}
\vspace{-14pt}
\end{table*}

\subsection{Dataset}

We utilize the English subset of RapBank to train the LM, which contains approximately $58,200$ songs with a total duration of 3,800 hours. After processing, we get the Basic, Standard and Premium subsets containing $1,321$, $295$ and $58$ hours of data respectively. We employ the entire RapBank to train the CFM model as it does not require any labels. 
We randomly reserved 200 samples for evaluation, with no singer overlapping with the training set. These samples are human-annotated to get the ground truth lyrics.

\subsection{Implementation and Hyperparameters}
We build a 6-layer LLaMA~\cite{llama} for lyrics-to-semantic modeling, with 116M parameters.
% During the training process, there is a 50\% chance of fully masking the accompaniment condition, as well as a 50\% chance of randomly masking the longest half of its tail. 
As mentioned earlier, to mitigate the train-inference mismatch of lengths in vocal-accompaniment pairs,
a masking strategy is applied probabilistically—there is a $50\%$ chance that the entire accompaniment condition will be masked, and for the other $50\%$ chance, a mask will be applied to a random length of the latter half of the accompaniment.
We first pre-train the LLaMA model on the Basic subset, followed by sequential supervised finetuning (SFT) on both the Standard and Premium subsets. We train the LLaMA model using $4$ NVIDIA V100 GPUs with a batch size of $16$ and gradient accumulation of $4$. The conditional flow matching model for semantic-to-spectrogram generation contains 129M parameters and is also trained using $4$ NVIDIA V100 GPUs. The batch size and gradient accumulation are $64$ and $4$, respectively. Each data segment is fixed at ten seconds in length, with shorter segments being padded and longer segments truncated. The sample steps is set to $20$.  For audio restoration, we employ the pre-trained BigVGAN-V2 $44.1$ kHz version\footnote{https://github.com/NVIDIA/BigVGAN}. The number of K-means clusters is set to $1024$, and the accompaniment feature shift $K$ is set to $150$ (3 secs).

\subsection{Evaluation Metrics}
%We conduct both subjective and objective evaluations on generated samples.
\subsubsection{Subjective Metrics}
We employ mean opinion score (MOS) with $95\%$ confidence intervals to evaluate multiple aspects of the generated rapping voice, including overall naturalness (MOS-N), singer similarity (MOS-S), vocal rhythmicity (MOS-R) and the stylistic and rhythmic alignment between vocals and acompaniment (MOS-M). The MOS scores are obtained by crowd-sourced listening tests, with 20 listeners involved. Each listener is instructed to evaluate the samples on a 5-point scale: 1 - bad, 2 - poor, 3 - fair, 4 - good, and 5 - excellent.
\subsubsection{Objective Metrics}
%Fr\'{e}chet
We employ word error rate (WER)  to measure intelligibility and speaker cosine similarity (SECS) to assess the timbre similarity, respectively. Specifically, we utilized the large-v3 version of Whisper\footnote{https://github.com/openai/whisper} to transcribe the generated rapping voice and calculate WER by comparing it to the human-annotated lyrics. For SECS measurement, we use WavLM-large fine-tuned on the speaker verification task\footnote{https://github.com/BytedanceSpeech/seed-tts-eval} to obtain speaker embeddings. These embeddings are then used to calculate the cosine similarity of generated rap against reference audio.
%Furthermore, we measure the overall quality of the rap using Fréchet Audio Distance (FAD). To evaluate the alignment between the vocal and the accompaniment, we employ Kullback–Leibler Divergence (KLD) and CLAP cosine similarity. Given that CLAP is trained using contrastive loss between audio and textual descriptions, we compute the CLAP similarity between the mixed audio and the accompaniment, as they are more likely to share the same semantic information.
Furthermore, we measure the overall quality of the rap using Fréchet Audio Distance (FAD) calculated using CLAP~\cite{clap}, and employ Kullback–Leibler Divergence (KLD) and CLAP cosine similarity~\cite{melodylm} to evaluate the distribution similairty and semantic correlation between real and generated rap.

\begin{table*}[!htbp]
\caption{Objective and Subjective evaluation of rapping vocals mixed with accompaniments generated by Freestyler and ablation models. "w/o Acco" means model trained without accompaniment. "w/o SFT" means remove supervised finetuning and "w/o Mask" means remove random masking on accompaniment condition.}
 \label{tab:mix}
  \centering
  \renewcommand{\arraystretch}{1.2}
\begin{tabular}{lccccccc}
\hline
Model               & MOS-N$\uparrow$ & MOS-M$\uparrow$ & FAD$\downarrow$ & KLD$\downarrow$ & CLAP$\uparrow$ & WER$\downarrow$ \\ \hline
GT                      & 4.35$\pm0.04$   & 4.21$\pm0.06$   & 0.07            & 0.01            & 0.90           & 31.7            \\
GT Rand Mix             & 3.23$\pm0.05$   & 3.06$\pm0.03$   & 0.16            & 0.33            & 0.45           & 31.7            \\ \hline
Freestyler              & 3.88$\pm0.07$   & 3.84$\pm0.06$   & 0.10            & 0.20            & 0.60           & 33.2            \\
\hspace{1em}w/o Acco    & 3.62$\pm0.06$   & 3.19$\pm0.04$   & 0.16            & 0.23            & 0.57           & 32.6            \\
\hspace{1em}w/o SFT     & 3.80$\pm0.05$   & 3.78$\pm0.07$   & 0.12            & 0.20            & 0.58           & 34.4            \\
\hspace{1em}w/o Mask    & 3.35$\pm0.05$   & 3.42$\pm0.05$   & 0.39            & 0.21            & 0.48           & 56.1            \\ \hline
\end{tabular}
\vspace{-8pt}
\end{table*}

\section{Evaluation Results}

\subsection{Vocal Quality Evaluation}
In this section, we assess the vocal generation capabilities of Freestyler, focusing on naturalness, rhythmicity, and intelligibility. We compare these metrics between the ground truth (GT) vocals and three distinct configurations of Freestyler. By employing random masking on accompaniment conditions during training, Freestyler is capable of operating with or without accompaniment during the inference stage. Furthermore, we conduct an ablation study to evaluate Freestyler in the absence of the accompaniment condition during both training and inference. The evaluation results are presented in Table~\ref{tab:vocal}.

The results demonstrate that Freestyler performs comparably to the GT across all metrics, highlighting its impressive capabilities in rapping voice generation. Notably, the naturalness and rhythmicity of the generated raps were significantly reduced when the accompaniment condition was excluded from both the training and inference stages, emphasizing the necessity of incorporating this condition. Furthermore, when the accompaniment is included during training but omitted during inference, the performance still exceeds that of scenarios where the accompaniment is excluded from both stages. This finding suggests that the model learns the rhythmic correlation between lyrics and the accompaniment when the latter is utilized in training, thereby enhancing performance even in lyrics-only inference.

Compared to normal speech, the WER across all models, including GT, is notably high, primarily due to Whisper's lack of training on singing data. The rapid tempo of rap, characterized by numerous conjunctions and varying intonations, presents significant challenges to Whisper's recognition accuracy. However, these models have similar WERs, suggesting that they achieve a level of intelligibility comparable to that of the ground truth.

\subsection{Accompaniment-mixed Rap Evaluation}
In this section, we further investigate rapping vocals accompanied by music. By utilizing segments from actual rap songs as the topline, we randomly select and mix unpaired vocals and accompaniment from our test set to serve as the bottomline. This setup allows us to examine the impact of stylistic and rhythmic mismatches on various metrics. As shown in Table~\ref{tab:mix}, the randomly mixed ground truth samples received the lowest scores, demonstrating the importance of style and rhythm alignment between vocals and accompaniment for natural rap performance. Conversely, Freestyler achieved scores comparable to GT, demonstrating its capability to generate natural-sounding raps. We also conducted experiments with three ablation systems: 1) removal of the accompaniment condition (w/o Acco); 2) removal of supervised finetuning (w/o SFT); and 3) removal of random masking (w/o Mask). The exclusion of the accompaniment condition resulted in a significant reduction in synchronization between vocals and accompaniment, leading to an overall decrease in perceived naturalness. The removal of SFT caused a slight decline in total metrics. Notably, the absence of random masking had the most pronounced effect, markedly reducing both naturalness and synchronization while causing a dramatic increase in word error rate (WER). This phenomenon arises because, in the absence of random masking, results in substantial mismatches between training and inferencing. If the lyrics do not conclude by the time the accompaniment ends, the model terminates prematurely; conversely, if the lyrics finish before the accompaniment is complete, the model generates nonsensical outputs, thus leading to the elevated WER metrics observed.

\subsection{Zero-shot Evaluation}
As previously discussed, the timbre within semantic tokens is incomplete. Therefore, we introduce reference audio conditions in both the lyrics-to-semantic and semantic-to-spectrogram stages. To assess the impact of reference audio conditions on speaker similarity across these stages, we examine three different combinations. The speaker similarity metrics presented in Table~\ref{tab:spk} demonstrate that: 1) Freestyler, which employs reference audio input in both stages, achieves a MOS-S score of 3.86 and a SECS score of 0.69, indicating a high degree of speaker similarity in the rapping voice generated by our proposed system; 2) The implementation of a single-stage timbre condition, whether in the first or second stage, results in a significant decline in similarity, thereby supporting our hypothesis that semantic tokens do not convey complete timbral information.

Additionally, we conducted experiments using speech instead of rap as the reference audio, enabling non-expert individuals to perform rap. As shown in the last row of the table, a satisfactory level of subjective similarity is achieved, illustrating Freestyler's remarkable generalization ability of zero-shot timbre control. However, the objective metric is low, probably due to the model we employed to compute SECS has never seen such a difference in style from the same speaker during training, resulting in low SECS score.

\begin{table}[]
\caption{Speaker similarity comparison results of different reference types with various condition methods.}
 \label{tab:spk}
  \centering
  \renewcommand{\arraystretch}{1.2}
\begin{tabular}{ccccc}
\hline
Ref Type & \begin{tabular}[c]{@{}c@{}}w/ LM\\ Cond\end{tabular} & \begin{tabular}[c]{@{}c@{}}w/ CFM\\ Cond\end{tabular} & MOS-S$\uparrow$                & SECS$\uparrow$                 \\ \hline
\multirow{3}{*}{Rap}      & \usym{2713}                  & \usym{2717}           & 3.44$\pm0.04$                 & 0.53                 \\
      & \usym{2717}                    & \usym{2713}         & 3.37$\pm0.05$                 & 0.50                 \\
      & \usym{2713}                    & \usym{2713}           & 3.86$\pm0.05$                 & 0.69                 \\ \hline
Speech   & \usym{2713}                    & \usym{2713}           & 3.64$\pm0.03$                 & 0.28                 \\ \hline
\end{tabular}
\vspace{-12pt}
\end{table}

\subsection{Vocal-Accompaniment Alignment Visualization}
To visualize the rhythmic correlation between the vocals and accompanying music rap songs, we analyzed both the ground truth and synthesized rap. As illustrated in Figure~\ref{fig:align}, the spectrograms of the ground truth accompaniment and vocals, along with the synthesized vocals, are presented sequentially. We manually annotated the positions of the beats in the accompaniment by drawing vertical lines across all three spectrograms. The results indicate that both the ground truth and generated vocals exhibit a strong correlation with the rhythm of the accompaniment, predominantly aligning with the beat positions. As the model does not have duration conditions and generates vocals freely, the synthesized vocals do not precisely match the ground truth. However, they demonstrate a similar pattern of alignment with the accompaniment. This finding underscores Freestyler's ability to utilize the accompaniment as a condition for generating rhythmically aligned rap.

\begin{figure}[ht]
\centering
\includegraphics[scale=0.55]{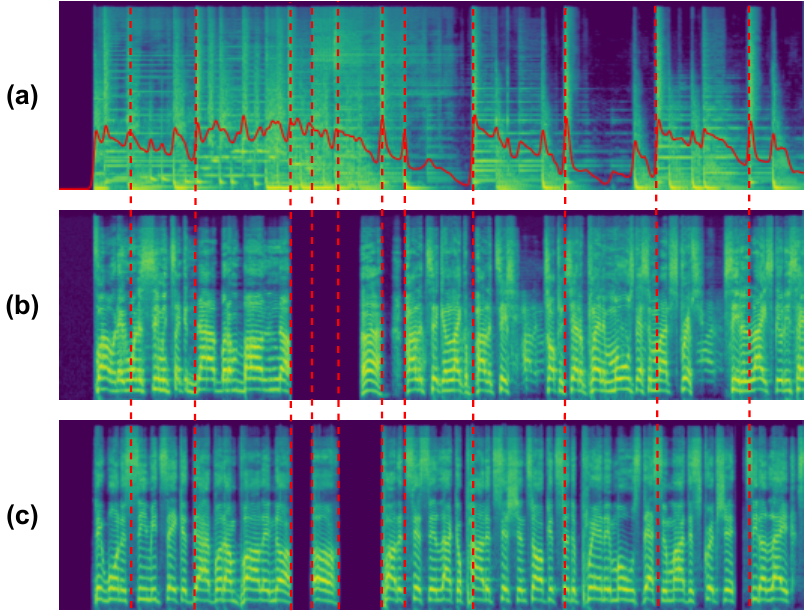}

\caption{The spectrogram of (a) GT accompaniment, (b) GT vocal and (c) Freestyler-generated vocal. Vertical lines are human-annotated beat positions in the accompaniment. The energy of the GT accompaniment is also drawn in (a).}
\label{fig:align}
\vspace{-8pt}
\end{figure}

\section{Conclusion}
In this paper, we propose Freestyler, the first rapping voice generation model that synthesizes high-quality rap vocals with enhanced naturalness and strong stylistic and rhythmic alignment with accompanying beats. 
% Freestyler divides the generation task into three stages: first, it utilizes lyrics and fine-grained accompaniment features as conditions to generate semantic tokens through a language model; second, it employs conditional flow matching to produce spectrograms; and finally, it leverages a vocoder to restore the audio.
Conditioned on fine-grained accompaniment features, Freestyler first generates semantic tokens through an autoregressive language model. Next, these tokens are converted to spectrograms using a conditional flow matching model, and the spectrograms are mapped to audios with a neural vocoder.
Additionally, we have developed an automatic pipeline to collect a large-scale rap dataset, which will be made publicly available along with the pipeline. Experimental results demonstrated the effectiveness of our proposed model structure and the strong rapping voice generation capability of Freestyler.

\bibliography{aaai25}
\newpage

\appendix

\begin{strip}
\centering
\textbf{\LARGE Appendix}
\end{strip}
%\titleformat{\section}[block]{\normalfont\Large\bfseries}{Appendix \thesection:}{1em}{}
\section{Dataset}
%86289 songs, avg duration
\subsection{Statistics}

\begin{figure}[h!]
\centering
\subfloat[Language]{
        \includegraphics[clip,scale=0.55]{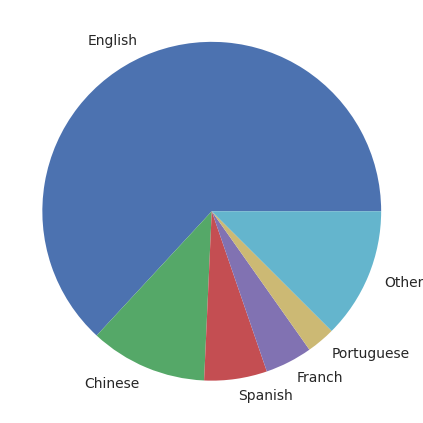}
        \label{fig:language}
    }
    \\
    \subfloat[Duration]{
        \includegraphics[clip,scale=0.35]{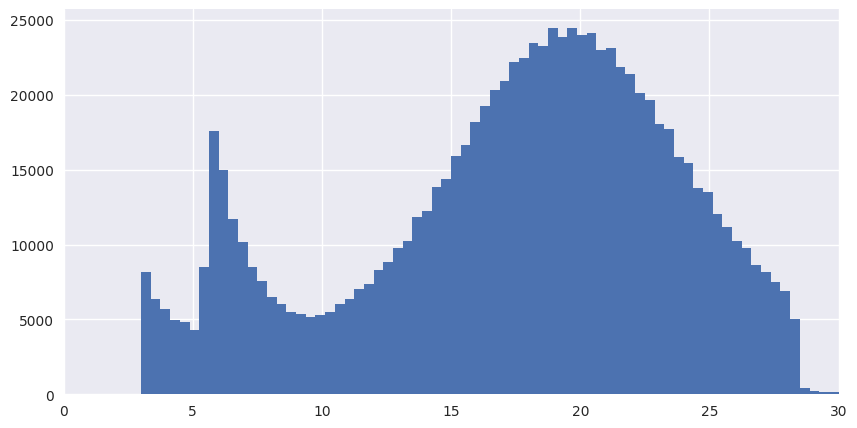}
        \label{fig:duration}
    }

\caption{The distribution of language and duration of RapBank}
\label{fig:align}
\vspace{-8pt}
\end{figure}

In this section, we present the statistics of RapBank. The RapBank dataset comprises links to a total of $94,164$ songs. However, due to the unavailability of certain videos, we successfully downloaded $92,371$ songs, amounting to $5,586$ hours of content, with an average duration of $218$ seconds per song. These songs span $84$ different languages. The Figure~\ref{fig:language} illustrates the top five languages based on the total hours of content. English has the highest duration, totaling 3,830 hours, which constitutes approximately two-thirds of the overall duration. We also utilize the English subset for training the language model in our experiments.

\begin{figure*}[htbp]
    \centering
    \subfloat[Phone-per-second]{%
        \includegraphics[width=0.32\textwidth]{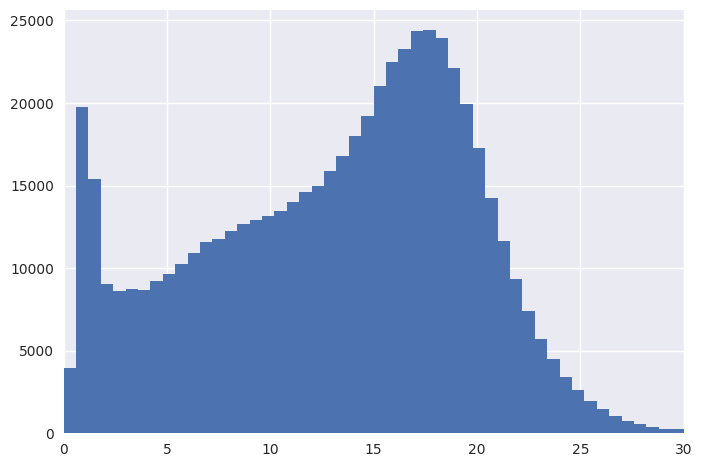} % 替换为你的图片文件名
        \label{fig:pps}
    }\hfill
    \subfloat[DNSMOS]{%
        \includegraphics[width=0.32\textwidth]{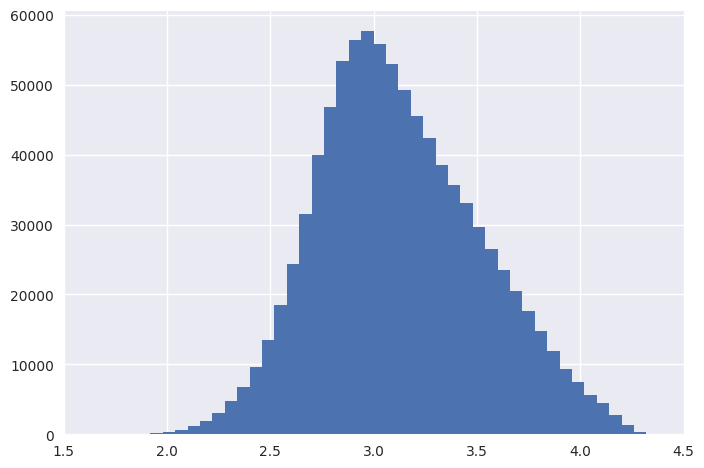} % 替换为你的图片文件名
        \label{fig:mos}
    }\hfill
    \subfloat[Primary Singer Percentage]{%
        \includegraphics[width=0.32\textwidth]{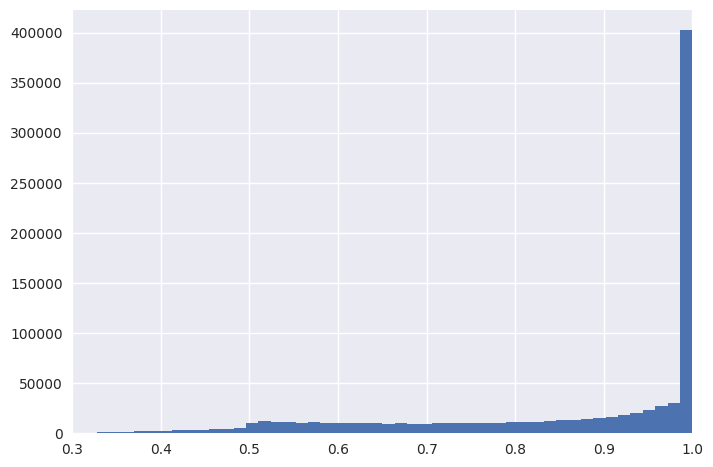} % 替换为你的图片文件名
        \label{fig:multi}
    }

    \caption{Quality Metrics}
    \label{fig:stat}
\vspace{-8pt}
\end{figure*}

As detailed in the main text, we merge the adjacent VAD-segmented data using a threshold sampled from a normal distribution with a mean of $18$ seconds. Segments with a duration of less than 3 seconds are discarded. As illustrated in Figure~\ref{fig:duration}, the durations of the resulting segments correspond with our expectation of a normal distribution with a mean value of $17.4$ seconds. After the separation and segmentation process, we obtained $904,548$ rap segments, totaling $4,353$ hours. Subsequently, we computed several quality-related metrics, including phone-per-second (PPS), DNSMOS, and the primary singer percentage. As depicted in Figure~\ref{fig:pps}, segments with lower PPS values contain an unusually high amount of data, which typically indicates hallucinations where the ASR model misrecognizes a long rap segment as a few words. In other cases, due to the fast tempo of rap, a lower PPS may mean that the segment is not rap. The DNSMOS scores presented in Figure~\ref{fig:mos} also conform to a normal distribution. Although the DNSMOS scores for rap are generally lower than those for speech, their relative values still provide a basis for evaluating rap quality. Regarding the multi-singer scenario, as shown in Figure~\ref{fig:multi}, most of the data did not feature multiple singers, and we filtered out segments with an insufficient primary singer percentage.

\begin{table}[]
\caption{Hyperparameters of Freestyler.}
\setlength{\tabcolsep}{3pt}
\label{tab:param}
\centering
\renewcommand{\arraystretch}{1.2}
\begin{tabular}{cccc}
\hline
Freestyler               & \multicolumn{2}{c}{Hyperparameter} & Params   \\ \hline
\multirow{5}{*}{Stage 1}       & Layers            & 6          & \multirow{5}{*}{116 M} \\
                                          & Hidden Dim        & 1024       &                        \\
                                          & Intermediate Dim  & 4096       &                        \\
                                          & Attention Heads   & 16         &                        \\
                                          & Speaker Dim       & 64         &                        \\ \hline
\multirow{6}{*}{Stage 2} & Down/Up Blocks & 2          & \multirow{6}{*}{129 M} \\
                                          & Mid Blocks        & 2          &                        \\
                                          & Input Dim         & 320        &                        \\
                                          & Intermediate Dim  & 768        &                        \\
                                          & Output Dim        & 128        &                        \\ 
                                          & Speaker Dim       & 64         &                        \\ \hline
\multirow{3}{*}{Stage 3}                  & Sampling Rate     & 44.1 kHz   & \multirow{3}{*}{122 M} \\
                                          & Upsampling Blocks  & 6        &                        \\
                                          & Upsampling Rate & 8, 4, 2, 2, 2, 2          &                        \\ \hline
\end{tabular}
\end{table}
\subsection{Subset}
We focused exclusively on the English language in RapBank for further analysis due to the high prevalence of hallucinations in lyrics from other languages. Processing of other languages will be addressed in future work. Utilizing the quality-related metrics described above, we established different thresholds to filter and categorize the data into three subsets of increasing quality: Basic, Standard, and Premium. As presented in Table~\ref{tab:data}, the three subsets have durations of $1,322$ hours, $295$ hours, and $58$ hours, respectively. The average segment duration across all three subsets is approximately $18$ seconds, indicating no significant correlation between duration and quality.

% 2+3
% basic 统计信息
% 数据划分

\begin{table*}[]
\caption{Thresholds for segmenting subsets and durations for RapBank and each subset}
 \label{tab:data}
  \centering
  \renewcommand{\arraystretch}{1.2}
\begin{tabular}{cccccc}
\hline
Subset     & \begin{tabular}[c]{@{}c@{}}DNSMOS\\ Threshold\end{tabular} & \begin{tabular}[c]{@{}c@{}}PPS\\ Threshold\end{tabular} & \begin{tabular}[c]{@{}c@{}}Primary Singer\\ Threshold\end{tabular} & Total Duration (h)  & Average Segment Duration (s) \\ \hline
Orig Songs & -                                                          & -                                                       & -                                                                  & 5586.2 & 227.7                        \\
RapBank    & -                                                          & -                                                       & -                                                                  & 4353.6 & 17.4                         \\
RapBank (English) & -                                                          & -                                                       & -                                                                  & 3830.1 & 17.3                        \\ \hline
Basic      & 2.5                                                        & 12-35                                                   & 0.8                                                                & 1322.0 & 18.5                         \\
Standard   & 3.5                                                        & 16-32                                                   & 0.9                                                                & 295.3  & 18.8                         \\
Premium    & 3.8                                                        & 18-30                                                   & 1.0                                                                & 58.3   & 18.7                         \\ \hline
\end{tabular}
\vspace{-8pt}
\end{table*}

\section{Model Archietechtures}
In this section, we provide the details of each of three stages of Freestyler. The architecture and hyperparameters are shown in Table~\ref{tab:param}.

\subsection{lyrics-to-semantic}
The backbone of lyrics-to-semantic is a LLaMA language model, featuring $6$ attention blocks. The hidden dimensions and intermediate dimensions are set to $1024$ and $4096$, respectively. It uses $16$ attention heads, each with $64$ dimensions. The reference encoder is composed of multiple linear layers with activation functions and concludes with an attention layer that does not use positional embeddings. An average pooling layer is applied along the time axis to obtain a global speaker embedding with 64 dimensions. 

\subsection{semantic-to-spectrogram}
We use a UNet-based conditional flow matching model for semantic-to-spectrogram modeling. The UNet architecture includes $6$ blocks: $2$ downsampling blocks, $2$ middle blocks, and $2$ upsampling blocks. The input dimension is $320$, which comprises the sum of mel channels multiplied by $2$ plus a 64-dimensional speaker embedding. The intermediate dimension is $768$, and the output dimension is $128$, corresponding to the mel dimension. The reference encoder has the same structure as the lyrics-to-semantic model but does not share parameters.

\subsection{spectrogram-to-audio}
We employ the $44.1$ kHz version of BigVGAN-V2 to restore audio from mel-spectrogram. The model comprises $6$ upsampling blocks, with upsampling rates multiplied to $512$, corresponding to $86.1$ frames per second. It does not require any fine-tuning to perform effectively on rap music as it is trained using datasets containing diverse audio types.

\end{document}